\documentclass[12pt,a4paper]{article}
\usepackage{lmodern}
\usepackage{here}
\usepackage{eurosym}
\usepackage{graphicx}
\usepackage{amsmath}
\usepackage{mathtools}
\usepackage{mathrsfs}
\usepackage{amssymb}
\usepackage{latexsym}
\usepackage{color}
\usepackage{epstopdf}
\usepackage{subcaption}
\usepackage{subcaption}
\usepackage[utf8]{inputenc} 
\usepackage{graphicx}
\usepackage{caption}
\usepackage{braket}
\usepackage[english]{babel}
\usepackage[T1]{fontenc}
\usepackage{float}
\usepackage{cite}
\usepackage[labelsep=space]{caption}

\usepackage{xcolor}
\usepackage{color}
\usepackage{hyperref}
\usepackage{cleveref}
\usepackage[switch]{lineno}
\usepackage[font=footnotesize,labelfont=bf]{caption}
\usepackage[hmargin=2cm,vmargin=2cm,rmargin=2cm, lmargin=2cm]{geometry}
\hypersetup{colorlinks=true,citecolor=red,filecolor=blue,linkcolor=blue,urlcolor=blue}
\DeclareUnicodeCharacter{FB01}{fi}
\DeclareUnicodeCharacter{200B}{}

\def\keywordname{{\bfseries \emph{Keywords}}}%
\def\keywords#1{\par\addvspace\medskipamount{\rightskip=0pt plus1cm
		\def\and{\ifhmode\unskip\nobreak\fi\ $\cdot$}\noindent\keywordname\enspace\ignorespaces#1\par}}

\begin{document}
	\renewcommand{\figurename}{\textbf{Fig.}}
	\renewcommand{\tablename}{\textbf{Table}}
	\newcommand{\un}{\mathds{1}}
	\newcommand{\1}{\mathbbm{1}}
	\newcommand{\vect}[1]{\boldsymbol{#1}}
	\newcommand{\era}{\end{array}}
\newcommand{\beq}{\begin{equation}}
	\newcommand{\eeq}{\end{equation}}
\newcommand{\beqar}{\begin{eqnarray}}
	\newcommand{\eeqar}{\end{eqnarray}}
\newcommand{\lb}{\label}
\thispagestyle{empty}
\baselineskip=18pt
\medskip
\begin{center}
	~~~~~~~~~~~~~~~
	\\
	\vspace{2cm}
	\noindent{{\textbf{Quantum Steering and Nonlocal Correlations Between Non-Interacting Delocalized Electrons Under Rashba Spin-Orbit Interaction}}}\\
	  
	\vspace{0.3cm}
	
	{\small Zakaria Bouafia}$^{a, }${\footnote{E-mail: \textsf{\href{mailto:zakaria.bouafia1@gmail.com}{zakaria.bouafia1@gmail.com}}}} and {\small Mostafa Mansour}$^{a, }${\footnote{E-mail: \textsf{\href{mailto:mostafa.mansour.fsac@gmail.com}{mostafa.mansour.fsac@gmail.com}}}} \\
	
	\vspace{0.3cm}
	
	\noindent $^{a}${{LMPHE, Department of Physics, Faculty of Sciences A\"{\i}n Chock,Hassan II University, Casablanca, Morocco}}
	
\end{center}

\vspace{1cm}

\begin{abstract}
	We investigate quantum steering and nonlocal correlations between two electrons in a two-dimensional electron gas (2DEG) as functions of Rashba spin-orbit interaction (RSOI) strength and inter-electron separation. We focus particularly on the Bi/Ag(111) system characterized by its strong RSOI ($\alpha_0 = 3.05\times10^{-11}$ eV~m), and we explore the influence of tuning intensity of RSOI and inter-electron distance on the dynamics of Bell nonlocality, uncertainty-induced nonlocality and steering. We find that, although increasing $\alpha_R$ initially suppresses quantum correlations, all three metrics exhibit a non-monotonic recovery as functions of $\alpha_R$, peaking near an optimal coupling strength $\alpha_R = 4.32\times10^{-11}$~eV~m across the range of inter-electron separations considered. This finding establishes RSOI as a critical control parameter for stabilizing quantum properties in two-dimensional electron gases against the decay of quantum correlations with inter-electron separation, and shows that the suppression and recovery of quantum resources within the Bi/Ag(111) system can be controlled by adjusting the inter-electron distance and carefully tuning the RSOI strength.
\end{abstract}

\vspace{0.3cm}

\keywords{Non-interacting delocalized electrons | Rashba spin-orbit interaction | Quantum steering | Bell nonlocality | Uncertainty-induced nonlocality}

\vspace{1cm}

\section{\label{sec:level1}Introduction}

Entanglement and discord-type quantum correlations are fundamental resources for the development of quantum information technologies. Quantum steering~\cite{einstein1935can, schrodinger1935discussion} goes a step beyond entanglement: it reflects the capacity of one party to steer the state of a remote system by performing only local measurements \cite{wiseman2007steering,uola2020quantum, du2021relationship,qin2017manipulating}, making it critical for device-independent protocols and fault-tolerant architectures~\cite{hsieh2016quantum, branciard2012one, he2015secure}. Quantum steering occupies a strictly intermediate position in the resource hierarchy: Bell nonlocality implies steerability, and steerability in turn implies entanglement; the reverse implications fail in general. \cite{hsieh2016quantum, uola2020quantum}. A hierarchy has also been demonstrated explicitly for two-qubit states under Markovian and non-Markovian noise\cite{costa2016generalized}.\\

The Rashba effect, resulting from spin-orbit coupling in inversion-asymmetric systems, has been extensively studied across a wide range of material platforms, including semiconductor quantum wells~\cite{koo2007appl, koo2009science, wunderlich2010science, chuang2015nature}, heavy-atom surfaces and alloys~\cite{miron2010natmat, miron2011nature}, oxide interfaces~\cite{gan2019prb, choe2019natcom, gao2018prl}, topological insulators~\cite{fan2014natmat, bahramy2012natcom}, low-dimensional nanomaterials~\cite{mourik2012science, gul2018natnano}, and spintronic devices \cite{Koo2020}. While entanglement dynamics in Rashba spin-orbit coupled systems are well-studied~\cite{aranzadi2023quantum, banouni2024non}, the interplay between Rashba interaction  strength ($\alpha_R$) and electron separation ($R$) in governing quantum steering, Bell nonlocality and uncertainty-induced nonlocality (UIN) remains unexplored to the best of our knowledge. Resolving this gap is therefore a prerequisite for engineering 2DEG-based devices in which $\alpha_R$ is tuned electrically to control nonlocal quantum correlations on demand.\\

In non-interacting systems, the exchange hole encodes intrinsic quantum correlations. Theoretical frameworks based on two-particle density matrices~\cite{dirac1931note, lowdin1955, vedral2003entanglement} reveal how RSOI reshapes these correlations in 2DEGs~\cite{aranzadi2023quantum, banouni2024non}. The two-fermion density matrix $\varrho_{AB}$ employed in this work follows the construction introduced by Aranzadi and Tamborenea~\cite{aranzadi2023quantum}, who used it to compute the exchange hole, concurrence, entanglement of formation, and quantum discord for the same 2DEG--Rashba system; this matrix was also employed in our earlier study of coherence and discord-type correlations~\cite{banouni2024non}. The present work is distinguished from both by its focus on three operationally distinct resources: Bell nonlocality, EPR steering, and uncertainty-induced nonlocality, none of which is a monotone of entanglement or discord and each of which certifies a different class of device-(in)dependent protocol. We further identify a non-monotonic recovery of all three resources at an optimal Rashba coupling $\alpha_R\approx4.32\times10^{-11}\,\mathrm{eV\,m}$, a regime not reported in Refs.~\cite{aranzadi2023quantum, banouni2024non}. This combination of resources has recently attracted attention across other spin-orbit-coupled platforms: Bell nonlocality together with quantum steering has been examined in Heitler–London-coupled spin systems~\cite{oumennana2025thermal} and in dephasing-controlled two-qubit spin systems~\cite{bougouffa2026magnetic}. The present work extends this line of inquiry to a delocalized, non-interacting 2DEG, where control is exercised through the Rashba coupling strength and the direct spatial separation between the electrons, rather than a localized qubit's coupling or detuning parameters, a distinction that is physically significant given the qualitatively different resource decay each control mechanism produces.\\ 

In 2DEGs, the RSOI can be tuned electrically via gate voltages, enabling controlled manipulation of spin textures~\cite{rashba1960properties, egues2002rashba}. This enables scalable, on-chip control of quantum phenomena, as evidenced by electric-field-driven Rashba splitting exceeding 100 meV in transition metal dichalcogenides~\cite{zhang2023voltage}. Such tunability positions 2DEGs as a versatile platform for spin-orbit qubits, particularly given their compatibility with semiconductor fabrication~\cite{burkard2023semiconductor}. Recent reviews identify 2DEGs as leading candidates for fault-tolerant quantum architectures, owing to their compatibility with industry-standard CMOS fabrication processes~\cite{burkard2023semiconductor, DeLeon2021}. This co-integration capability, demonstrated in silicon-based 2DEG platforms~\cite{Veldhorst2017, Zwanenburg2013}, enables scalable manufacturing of spin-based qubit arrays with classical control circuitry. In 2DEGs, with spin-orbit interaction (SOI), delocalization between electrons sustains nonclassical correlations, such as entanglement and quantum discord, that decay gradually with spatial separation but that can be amplified by increasing the SOI strength \cite{aranzadi2023quantum, banouni2024non}. Material advances, including InGaAs/InAlAs heterostructures (coherence times exceeding 100 ns) and CMOS-compatible 2DEGs, have highlighted the role of delocalization in fault-tolerant architectures \cite{Veldhorst2017, Mi2023}. In this work, we investigate how the interplay between RSOI strength $\alpha_R$ and inter-electron separation $R$ governs Bell nonlocality, UIN, and quantum steering. We focus on the Bi/Ag(111) system, taking $\alpha_0 = 3.05\times10^{-11}$~eV~m as adopted in Ref.~\cite{aranzadi2023quantum} from the giant Rashba splitting measured in this surface alloy by Ref.~\cite{ast2007giant}. This system exhibits exceptionally strong RSOI due to surface alloying and structural corrugation \cite{aranzadi2023quantum, ast2007giant}. \\

The paper is organized as follows. Section~\ref{sec:level2} reviews the quantum resources used in this study (quantum steering, Bell nonlocality, and uncertainty-induced nonlocality). In Section \ref{sec:level3}, we introduce the 2DEG Hamiltonian and we give the corresponding density matrix. In Section \ref{sec:level4}, we discuss the results of the behavior of the key metrics against RSOI, and the separation between electrons. Section \ref{sec:level5} summarizes the main conclusions and gives some implications of our findings.

\section{\label{sec:level2}Quantum Metrics}

This section introduces three quantum metrics used in this study: Bell nonlocality, uncertainty-induced nonlocality, and quantum steering. Their purpose is to characterize the correlations of delocalized electrons in a 2DEG in the presence of RSOI. All three are computed analytically from the two-fermion density matrix of the 2DEG and studied as functions of $\alpha_R$ and $R$.
\subsection{\label{subsec:level21}Bell Nonlocality}
Bell nonlocality provides a well-established framework for examining nonlocal correlations in bipartite quantum systems. In this context, the CHSH inequality serves as one of the most straightforward and frequently employed inequalities to identify Bell nonlocality violations for any two-party state $\varrho_{AB}$ \cite{clauser1969proposed, bartkiewicz2013entanglement, horodecki1995violating}. Any local hidden-variable model restricts the correlations between subsystems $A$ and $B$ to the classical bound given by the inequality:
\begin{equation}
	|\operatorname{Tr}\left(\varrho_{AB}\,\mathcal{B}_{\mathrm{CHSH}}\right)| \leq 2,
\end{equation}
which constrains the strength of classical correlations. The Bell operator ($\mathcal{B}_{\mathrm{CHSH}}$), which represents the fundamental element for the CHSH inequality, is defined as follows:
\begin{align}
	\mathcal{B}_{\mathrm{CHSH}}&=\boldsymbol{\alpha}\cdot\boldsymbol{\sigma}\otimes\left(\boldsymbol{\beta}+\boldsymbol{\beta}^{\prime}\right)\cdot\boldsymbol{\sigma}\nonumber\\
	&+\boldsymbol{\alpha}^{\prime}\cdot\boldsymbol{\sigma}\otimes\left(\boldsymbol{\beta}-\boldsymbol{\beta}^{\prime}\right)\cdot\boldsymbol{\sigma},
\end{align}
here, the vectors $\boldsymbol{\alpha}\left(\boldsymbol{\alpha}^{\prime}\right)$ and $\boldsymbol{\beta}\left(\boldsymbol{\beta}^{\prime}\right)$ represent unit vectors corresponding to the measurement settings on two parties $A$ and $B$, respectively. $\boldsymbol{\sigma}=\left(\sigma_x, \sigma_y, \sigma_z\right)$ represents the two-dimensional usual Pauli vector. \\
Maximizing over $\boldsymbol{\alpha}$, $\boldsymbol{\alpha}^{\prime}$, $\boldsymbol{\beta}$, and $\boldsymbol{\beta}^{\prime}$ gives the largest attainable value of the Bell operator for the state $\varrho_{AB}$:
\begin{align}
	\mathcal{\hat{B}}_{\mathbf{CHSH}}(\varrho_{AB})&=\max _{\mathcal{B}_{\mathrm{CHSH}}}\left|\operatorname{Tr}\left(\varrho_{AB}\, \mathcal{B}_{\mathrm{CHSH}}\right)\right| \nonumber\\
	&=2\sqrt{\mathcal{M}(\varrho_{AB})},
\end{align}
where $\mathcal{M}(\varrho_{AB})=\max _{l<k}\left\{\omega_l+\omega_k\right\} \leq 2$ and $\omega_{l_{(l=1,2,3)}}$ represents the eigenvalues of $\mathcal{W}=C^{T}C$, a correlation matrix, whose elements $C_{ij}=\operatorname{Tr}\left[\varrho_{AB}\left(\sigma_i \otimes \sigma_j\right)\right]$. When $\mathcal{M}(\varrho_{AB})>1$, the CHSH inequality is violated \cite{horodecki1995violating}. In the present investigation, the degree of CHSH inequality violation will be measured through the Bell nonlocality metric $\mathcal{B}(\varrho_{AB})$. $\mathcal{B}(\varrho_{AB})$ writes as follows:
\begin{equation}
	\mathcal{B}(\varrho_{AB}) = \sqrt{\max(0, \mathcal{M}(\varrho_{AB})-1)}.\label{B}
\end{equation}
This normalizes $\mathcal{B}(\varrho_{AB})$ against the Tsirelson bound $\mathcal{M}=2$, consistent with the normalization of $\mathcal{S}(\varrho_{AB})$ in Eq.~\eqref{Sab}. Here, $0\leq\mathcal{B}(\varrho_{AB})\leq1$, where $\mathcal{B}(\varrho_{AB})=0$ indicates no CHSH inequality violation, $\mathcal{B}(\varrho_{AB})>0$ refers to growing violations, and $\mathcal{B}(\varrho_{AB})=1$ indicates the maximum violation.
\subsection{\label{subsec:level22}Uncertainty-Induced Nonlocality}
The concept of UIN, introduced in \cite{wu2014uncertainty}, offers a different perspective on nonlocal correlations, distinct from measurement-induced nonlocality (MIN) \cite{luo2011measurement, hu2015measurement}. UIN has recently been used to probe thermal and gravitational decoherence of quantum correlations in gravitational cat states \cite{dahbi2023skew}.  UIN is a refined measure of MIN, for $\varrho_{AB}$  as \cite{wu2014uncertainty}:
\begin{equation}\label{Uc}
	\mathcal{U}_{c}(\varrho_{AB})=\max_{\zeta^S}\mathcal{I}\left(\varrho_{AB}, \zeta^S\right),
\end{equation}
where the skew information $\mathcal{I}(\varrho_{AB}, \zeta)$ is given by \cite{wigner1963information, luo2003wigner}:
\begin{equation}
	\mathcal{I}(\varrho_{AB}, \zeta)=-\operatorname{Tr}[\varrho_{AB}^{1/2}, \zeta]^2/2.
\end{equation}
The uncertainty introduced by the observable $\zeta$ acting on the two-qubit system $\varrho_{AB}$ is quantified by the skew information. The maximization in Eq.~(\ref{Uc}) runs over all commuting observables that are locally maximally informative, $\zeta^S=\zeta_A^S\otimes\mathbb{I}_B$, with $\mathbb{I}_B$ the identity operator on subsystem $B$ and $\zeta_A^S$ a Hermitian operator on subsystem $A$ with non-degenerate eigenvalues. The explicit formula for the uncertainty-induced nonlocality of a given quantum state $\varrho_{AB}$ in a $(2 \otimes d)$-dimensional space is thoroughly explained in \cite{wu2014uncertainty}. It is defined as:
\begin{align}\label{U}
	\mathcal{U}_{c}(\varrho_{AB})=\left\{\begin{array}{cl}
		&1-n_{\min }(\mathcal{N}),\quad\boldsymbol{v}=\mathbf{0}; \\
		&1-\frac{1}{|\boldsymbol{v}|^2}\boldsymbol{v}\mathcal{N}\boldsymbol{v}^T,\quad\boldsymbol{v}\neq\mathbf{0},
	\end{array}\right.
\end{align}
Here, $\boldsymbol{v}$ is the Bloch vector of subsystem $A$, with $\boldsymbol{v}^T$ its transpose. $n_{\min}(\mathcal{N})$ is the smallest eigenvalue of the symmetric real matrix $\mathcal{N}_{3\times3}$, whose elements are defined as
\begin{equation}
	(\mathcal{N})_{ij}=\operatorname{Tr}\left\{\sqrt{\varrho_{AB}}\left(\sigma_i^A \otimes \mathbb{I}_B\right) \sqrt{\varrho_{AB}}\left(\sigma_j^A \otimes \mathbb{I}_B\right)\right\},
\end{equation}
with $\{\sigma_{i(j)}^A\}_{i(j)=x,y,z}$ the standard Pauli matrices acting on subsystem $A$.
\subsection{\label{subsec:level23}Quantum Steering}
For  the quantum state $\varrho_{AB}$, quantum steering is evaluated by employing the Cavalcanti-Jones-Wiseman-Reid (CJWR) inequality \cite{cavalcanti2009experimental}:
\begin{equation}\label{eq21}
	\mathcal{F}(\varrho_{AB})=\frac{1}{\sqrt{3}}\sum_{j=1}^3\operatorname{Tr}(\varrho_{AB}A_jB_j) \leq 1,
\end{equation}
where $A_j$ and $B_j$ are projectors built from the Pauli operators, and a violation of Eq.~(\ref{eq21}) signals that the system exhibits quantum steering. Optimizing the left-hand side of (Eq. (\ref{eq21}))  over the measurement directions $A_j, B_j$ yields the steering function
\begin{equation}
	\mathcal{F}(\varrho_{AB})=\sqrt{\sum_{i=1}^3m_i^2},
\end{equation}
Here, $m_i$ ($i\in\{1,2,3\}$) denote the eigenvalues of $\sqrt{T^{\top}T}$, arranged in decreasing order, with $T$ having elements $T_{(i,j)}=\operatorname{Tr}[\varrho_{AB}(\sigma_i\otimes\sigma_j)]$. The steering quantifier $\mathcal{S}(\varrho_{AB})$ is then given by \cite{costa2016quantification}:
\begin{equation}\label{Sab}
	\mathcal{S}(\varrho_{AB})=\max\left(0, \frac{\mathcal{F}(\varrho_{AB})-1}{\sqrt{3}-1}\right).
\end{equation}
Quantum steering is useful in particular for one-sided device-independent protocols, and fault-tolerant architectures, since only one party needs to trust its measurement devices.

\section{\label{sec:level3}Two-Fermion Spin System in a 2DEG}

We consider a 2DEG confined to the xy-plane, in which RSOI emerges from structural inversion asymmetry (SIA). To isolate RSOI-driven effects, we exclude Coulomb interactions and focus our analysis on spin-texture structure and exchange-hole suppression. The non-interacting single-particle Hamiltonian governing this system is expressed as \cite{bychkov1984oscillatory}:
\begin{align}\label{HR}
	\hat{H} = \frac{e \hbar^2}{(2m_0 c)^2} ({\vec{k}} \times {\vec{E}}) \cdot \boldsymbol{\vec{\sigma}} - \frac{\hbar^2}{2m^*} \hat{\nabla}^2,
\end{align}
where the Rashba spin-orbit interaction Hamiltonian is given by $H_{\text{SO}} = \frac{e \hbar^2}{(2m_0 c)^2} ({\vec{k}} \times {\vec{E}}) \cdot \boldsymbol{\vec{\sigma}}$, with $m_0$ is the free electron mass, $\vec{k} = (k_x, k_y, 0)$ is the wavevector in the plane of the 2DEG, ${\vec{E}} = (0, 0, E_z)$ is the electric field perpendicular to the 2DEG, with $\boldsymbol{\vec{\sigma}} = (\sigma_x, \sigma_y, \sigma_z)$ denoting the Pauli spin matrices. For an electric field in the $z$-direction, the cross product simplifies to ${\vec{k}} \times {\vec{E}} = (k_y, -k_x, 0) E_z$. Thus, the Rashba Hamiltonian becomes $H_{\text{SO}} = \alpha (- k_x \sigma_y + k_y \sigma_x)$ and reflects spin-momentum locking, a hallmark of inversion-asymmetric systems \cite{Manchon2015}. Here $\alpha \propto E_z$ is the Rashba spin-orbit coupling parameter. The parameter $ m^* $ is the effective mass of the fermions in the 2DEG, and $ \hat{\nabla}^2 = \frac{\partial^2}{\partial x^2} + \frac{\partial^2}{\partial y^2} $ denotes the Laplacian operator. Solving this Hamiltonian, we find two spin-split energy branches, labeled by the spin-splitting parameter $ \gamma = \pm 1 $, associated with the distinct eigenstates. The corresponding eigenenergies  for a given momentum $ \vec{k}  $ are \cite{bychkov1984oscillatory}:
\begin{equation}\label{eq13}
	E_\pm(\mathbf{k}) = \frac{\hbar^2 k^2}{2 m^*} \pm k \alpha.
\end{equation}
The kinetic term $ \frac{\hbar^2 k^2}{2m^*} $ represents the particle's energy without spin-orbit interaction, while the $ \pm \alpha k $ term, which is linear in $ k $, accounts for the spin-dependent energy shift induced by the RSOI. The corresponding eigenfunctions are given by:
\begin{equation}
	\psi_{\pm,\mathbf{k}}(\mathbf{r}) = \frac{e^{i \mathbf{k} \cdot \mathbf{r}}}{\sqrt{2 A}} \chi^{\pm}(\phi) = \frac{e^{i \mathbf{k} \cdot \mathbf{r}}}{\sqrt{2 A}} \begin{pmatrix} \pm i e^{-i \phi} \\ 1 \end{pmatrix},
\end{equation}
where $ \mathbf{r} $ is the position vector, $ A $ is the area of the system, and $ \phi $ is the polar angle of $ \vec{k} $ in momentum space. The spin components of the wave functions $ \chi^{\pm}(\phi) $ are orthogonal, i.e., $ \langle \chi^{\pm} | \chi^{\mp} \rangle = 0 $, which implies that the electron spins lie within the $ xy $-plane and point in opposite directions, forming a pair of oppositely oriented spins in this plane. As a result, the spinor structure of $\psi_{\pm,k}(r)$ reveals that the spin orientation is coupled to the momentum direction $\mathbf k$. Explicitly, for $\chi^+(\phi)\propto(ie^{-i\phi},1)^T$ one finds $\langle\sigma_x\rangle=\sin\phi$, $\langle\sigma_y\rangle=-\cos\phi$, $\langle\sigma_z\rangle=0$, so the physical spin direction $(\langle\sigma_x\rangle,\langle\sigma_y\rangle,\langle\sigma_z\rangle)$ is a unit vector confined to the $xy$-plane and rotated $-90^\circ$ from $\hat{\mathbf k}$, i.e., locked perpendicular to the momentum, consistent with the known helical spin texture of the Rashba interaction. The $-$ branch gives the opposite in-plane direction, consistent with the orthogonality $\langle\chi^+|\chi^-\rangle=0$ noted above. This momentum-spin locking is a direct consequence of the RSOI, which breaks spin degeneracy and causes the energy spectrum to split into two distinct branches. For analyzing the system with multiple fermions, it is suitable to employ the second quantization formalism. The many-fermion Hamiltonian for a 2DEG under RSOI in this formalism becomes:
\begin{equation}\label{HR0}
	\hat{H} = \sum_{k ,\gamma} E_\gamma(k ) \hat{c}_{k \gamma}^\dagger \hat{c}_{k \gamma},
\end{equation}
where ($\hat{c}_{k\gamma}^\dagger$) ($\hat{c}_{k\gamma}$) is the fermion creation (annihilation) operator for states labeled by ($k$) and ($\gamma$). Here, we explore the situation in which the density of electrons in the 2DEG is considerably high ($n>\frac{m^{*2}\alpha^2}{\pi\hbar^4}$), allowing that both energy branches are occupied by electrons. In this scenario, the many-body ground state of the Hamiltonian (Eq. (\ref{HR0})) is noted by \cite{bruus2004many}:
\begin{equation} \label{eq16}
	|\Phi_0\rangle=\left(\prod_{|k|}^{k_F^+}\hat{c}_{k+}^{\dagger}\prod_{|k|}^{k_F^-}\hat{c}_{k-}^{\dagger}\right)|0\rangle,
\end{equation}
with, $|0\rangle$ is the corresponding vacuum state. The ground state $|\Phi_0\rangle$ corresponds to two filled Fermi seas with Fermi momenta $k_F^+$ and $k_F^-$, the radii of two concentric Fermi circles in the 2D Brillouin zone. The Fermi momenta $k_F^\pm$ of the two spin-split branches are determined by imposing a common Fermi energy across both branches [Eq.~(13)],
$$
\frac{\hbar^2(k_F^+)^2}{2m^*}+\alpha_R k_F^+ = \frac{\hbar^2(k_F^-)^2}{2m^*}-\alpha_R k_F^-,
$$
together with the density constraint $(k_F^+)^2+(k_F^-)^2=4\pi n$. Defining the Rashba wavevector $k_{so}\equiv m^*\alpha_R/\hbar^2$, these two conditions give $k_F^- - k_F^+ = 2k_{so}$, and combined with the density constraint yield the explicit expressions
\begin{equation}
	k_F^\pm = \sqrt{2\pi n - k_{so}^2}\ \mp\ k_{so}, 
\end{equation}
consistent with the general two-subband Fermi wavevector formula derived for Rashba–Dresselhaus 2DEGs in the pure-Rashba limit \cite{Maytorena2006}. The high-density condition $n > k_{so}^2/\pi$ (equivalently $n>m^{*2}\alpha_R^2/(\pi\hbar^4)$) ensures $k_F^+>0$, i.e., that both branches are occupied, consistent with the ground state~(\ref{eq16}). To analyze the two-fermion density matrix of a 2DEG under the RSOI, we define its matrix elements, including a normalization factor, as follows \cite{vedral2003entanglement}:
\begin{equation}\label{var}
	\varrho_{ss';tt'}=\langle \Phi_0|\mathcal{\hat{\psi}}_{t'}^{\dagger}(\mathbf{\mathbf{r'}})\mathcal{\hat{\psi}}_{t}^{\dagger}(\mathbf{\mathbf{r}})\mathcal{\hat{\psi}}_{s}(\mathbf{r})\mathcal{\hat{\psi}}_{s'}(\mathbf{r'})|\Phi_0\rangle.
\end{equation}
Here, the field operator $\hat{\psi}^{\dagger}_s(\mathbf{r})$ is the creation operator for a fermion at position $\mathbf{r}$, with a spin projection $s$ aligned along the $x$-direction. The expression for $\hat{\psi}_s(\mathbf{r})$ is given as \cite{aranzadi2023quantum}:
\begin{equation}\label{p}
	\hat{\psi}_s(\mathbf{r})=\sum_{k,\gamma'}e^{i k\mathbf{r}}\frac{1}{2\sqrt{2A}}\left(\gamma'ie^{i\Phi_k}+s\right)\hat{c}_{k \gamma'}.
\end{equation}
By substituting $\hat{\psi}_s(\mathbf{r})$, as defined in Eq. (\ref{p}), into Eq. (\ref{var}), we obtain the following result \cite{vedral2003entanglement, banouni2024non}:
\begin{align}
	\varrho_{ss';tt'}=&\sum_{\mathbf{k}\gamma}\sum_{\mathbf{k'}\gamma'}\sum_{\mathbf{l}\sigma}\sum_{\mathbf{l'}\sigma'}\frac{e^{-i(\mathbf{k}-\mathbf{l})\mathbf{r}}e^{-i(\mathbf{k'}-\mathbf{l'})\mathbf{r'}}}{2^6A^2}\nonumber\\
	&\times(t'-\gamma ie^{i\Phi_\mathbf{k}})(t-\gamma' ie^{i\Phi_\mathbf{k'}}) \nonumber\\
	&\times(s'+\sigma ie^{-i\Phi_\mathbf{l}})(s+\sigma' ie^{-i\Phi_{\mathbf{-l'}}})\nonumber\\
	&\times(\delta_{\mathbf{k}\mathbf{l}}\delta_{\mathbf{k'}\mathbf{l'}}\delta_{\gamma\sigma}\delta_{\gamma'\sigma'}-\delta_{\mathbf{k}\mathbf{l'}} \delta_{\mathbf{k'}\mathbf{l}}\delta_{\gamma\sigma'}\delta_{\gamma'\sigma}).
\end{align}
Applying the normalization condition $\operatorname{Tr}(\varrho)=1$, we work in the computational basis $\{|\uparrow\uparrow\rangle, |\uparrow\downarrow\rangle, |\downarrow\uparrow\rangle, |\downarrow\downarrow\rangle\}$, with the spin quantization axis along $\hat{\mathbf{x}}$. The resulting two-spin density matrix $ \varrho_{AB} $ for the two-fermion system, first derived in \cite{aranzadi2023quantum} is:
\begin{equation}
	\label{eq:rashba_matrix}
	\varrho_{AB} = \frac{1}{4 - 2\Upsilon}
	\begin{pmatrix}
		1-\Gamma_1^2 & \Gamma_1\Gamma_2 & -\Gamma_1\Gamma_2 & \Gamma_2^2 \\
		\Gamma_1\Gamma_2 & 1-\Gamma_2^2 & -\Gamma_1^2 & \Gamma_1\Gamma_2 \\
		-\Gamma_1\Gamma_2 & -\Gamma_1^2 & 1-\Gamma_2^2 & -\Gamma_1\Gamma_2 \\
		\Gamma_2^2 & \Gamma_1\Gamma_2 & -\Gamma_1\Gamma_2 & 1-\Gamma_1^2
	\end{pmatrix},
\end{equation}
where $\Upsilon=\Gamma_{1}^2+\Gamma_{2}^2$, with $\Gamma_1$ and $\Gamma_2$ given by \cite{aranzadi2023quantum}:
\begin{align}
	\Gamma_1=&\frac{2[k_F^+J_1(k_F^+Rr_0)+k_F^-J_1(k_F^-Rr_0)]}{(k_F^{+2}+k_F^{-2})Rr_0},\label{go}\\
	\Gamma_2=&\frac{\pi\, k_F^+\bigl(H_0(k_F^+ Rr_0)J_1(k_F^+ Rr_0) - H_1(k_F^+ Rr_0)J_0(k_F^+ Rr_0)\bigr)}{(k_F^{+2}+k_F^{-2})Rr_0}\nonumber\\
	&- \frac{\pi\, k_F^-\bigl(H_0(k_F^- Rr_0)J_1(k_F^- Rr_0) - H_1(k_F^- Rr_0)J_0(k_F^- Rr_0)\bigr)}{(k_F^{+2}+k_F^{-2})Rr_0}, \label{ga}
\end{align}
where, $\Gamma_1$ quantifies same-spin Pauli exclusion and $\Gamma_2$ captures RSOI-driven antiparallel spin correlations. The parameter $R= \frac{|r-r'|}{r_0}$ is the inter-fermion separation in the 2DEG with $r_0=\left(\sqrt{\pi n}\right)^{-1}$ is the associated characteristic radius. The terms $J_n$ and $H_n$ refer to Bessel and Struve functions, respectively. The density matrix $ \varrho_{AB} $ represents a pair of spin-$ \frac{1}{2} $ degrees of freedom separated by a distance $ R $. The reduced state of subsystem $A$ is obtained as:
\begin{equation}
	\varrho_A = \mathrm{Tr}_B(\varrho_{AB}) = \frac{1}{4-2\Upsilon}
	\begin{pmatrix} 2-\Upsilon & 0 \\ 0 & 2-\Upsilon \end{pmatrix} = \frac{1}{2}\,\mathbb{I}_2,
\end{equation}
which is maximally mixed for all values of $\alpha_R$ and $R$, confirming that the Bloch vector $\mathbf{v}$ of subsystem $A$ vanishes identically.\\
The eigenvalues and eigenvectors of $ \varrho_{AB} $ are given by:
	\begin{align}
		\label{eq1} &E_1=\frac{\Upsilon+1}{4-2\Upsilon},& \quad |U_1\rangle&=\frac{1}{\sqrt{2+2\Big(\big|\frac{\Gamma_1}{\Gamma_2}\big|\Big)^2}}|00\rangle+\frac{\Gamma_1}{\Gamma_2\sqrt{2+2\Big(\big|\frac{\Gamma_1}{\Gamma_2}\big|\Big)^2}}|01\rangle \nonumber\\
		& & &-\frac{\Gamma_1}{\Gamma_2\sqrt{2+2\Big(\big|\frac{\Gamma_1}{\Gamma_2}\big|\Big)^2}}|10\rangle+\frac{1}{\sqrt{2+2\Big(\big|\frac{\Gamma_1}{\Gamma_2}\big|\Big)^2}}|11\rangle,\\
		\label{eq2} &E_2=\frac{\Upsilon-1}{2\Upsilon-4},& \quad |U_2\rangle&=-\frac{1}{\sqrt{2}}|00\rangle+\frac{1}{\sqrt{2}}|11\rangle,\\
		\label{eq3} &E_3=\frac{\Upsilon-1}{2\Upsilon-4},& \quad |U_3\rangle&=\frac{\Gamma_1}{\Gamma_2\sqrt{1+\Big(\big|\frac{\Gamma_1}{\Gamma_2}\big|\Big)^2}}|00\rangle+\frac{1}{\sqrt{1+\Big(\big|\frac{\Gamma_1}{\Gamma_2}\big|\Big)^2}}|10\rangle,\\
		\label{eq4} &E_4=\frac{\Upsilon-1}{2\Upsilon-4},& \quad |U_4\rangle&=-\frac{\Gamma_1}{\Gamma_2\sqrt{1+\Big(\big|\frac{\Gamma_1}{\Gamma_2}\big|\Big)^2}}|00\rangle+\frac{1}{\sqrt{1+\Big(\big|\frac{\Gamma_1}{\Gamma_2}\big|\Big)^2}}|01\rangle.
	\end{align}
Using the elements of $\varrho_{AB}$ [Eq. (\ref{eq:rashba_matrix})], we derive explicit expressions for the quantifiers used in this study. Substituting the correlation matrix of $\varrho_{AB}$ into Eq. (\ref{B}) yields:
\begin{equation}\label{eq31}
	\mathcal{B}(\varrho_{AB}) = \sqrt{\max(0,\frac{2\Upsilon^2}{(2-\Upsilon)^2}-1)}.
\end{equation}
The uncertainty-induced nonlocality $U_c(\varrho_{AB})$ is obtained by substituting the eigenvalues of $\mathbf{N}$ into [Eq. (\ref{U})]  with $\mathbf{v}=0$, yielding:
\begin{equation}
	U_c(\varrho_{AB}) = 1 - \frac{(1-\Upsilon) + \sqrt{1-\Upsilon^2}}{2 - \Upsilon}, \label{eq:UIN} 
\end{equation}
where $\Upsilon = \Gamma_1^2 + \Gamma_2^2 \in [0,1]$. This expression is manifestly real-valued and satisfies $U_c(\varrho_{AB}) \in [0,1]$ throughout the physical parameter range. The quantum steering $\mathcal{S}(\varrho_{AB})$ Eq. (\ref{Sab}) is found as:
\begin{equation}\label{eq33}
	\mathcal{S}(\varrho_{AB})=\max \left(0, \frac{\sqrt{\frac{3\Upsilon^2}{(2- \Upsilon)^2}}-1}{\sqrt{3}-1}\right).
\end{equation}
In the following section, we will utilize the previously derived results to investigate the dynamical behavior of three metrics, considering the influence of the Rashba coupling and the spatial separation between two non-interacting delocalized electrons in a 2DEG system.

\section{\label{sec:level4}Results and Discussion}

\begin{figure*}[t]
	\centering
	\subfloat[]{%
		\includegraphics[width=.31\textwidth]{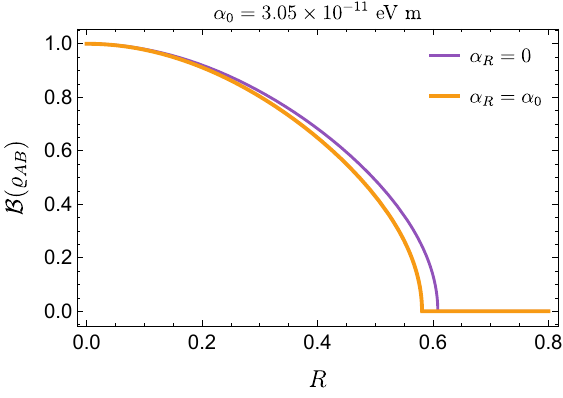}
		\label{1a}}
	\hfill
	\subfloat[]{%
		\includegraphics[width=.31\textwidth]{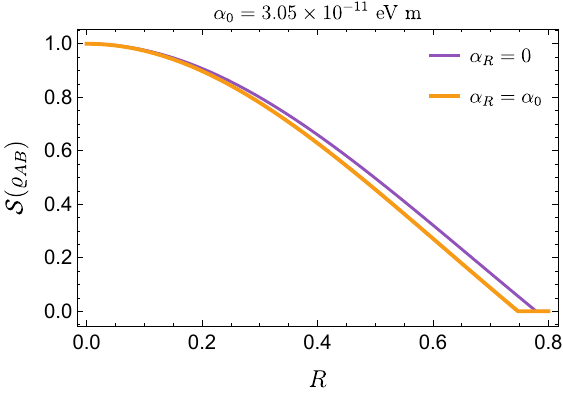}
		\label{1b}}
	\hfill
	\subfloat[]{%
		\includegraphics[width=.31\textwidth]{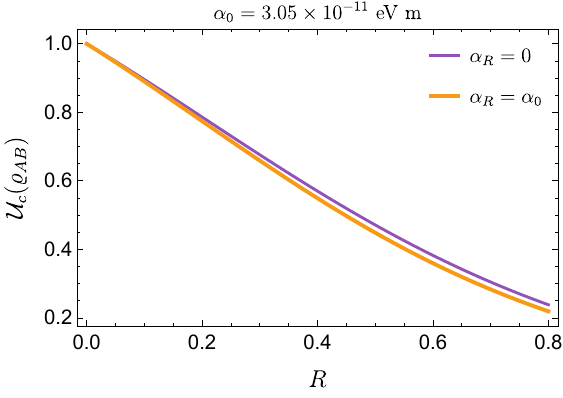}
		\label{1c}}\\
	\subfloat[]{%
		\includegraphics[width=.31\textwidth]{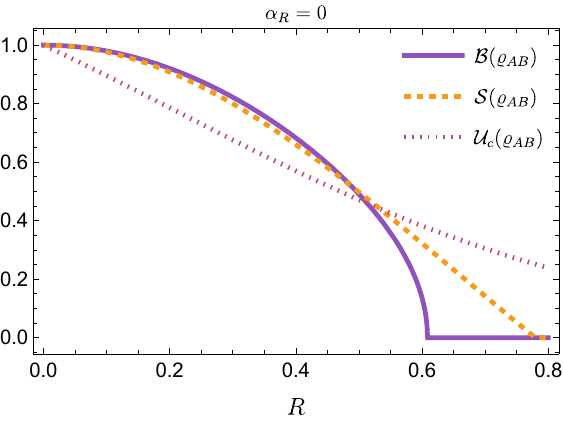}
		\label{1d}}
	\hspace{0.7cm}
	\subfloat[]{%
		\includegraphics[width=.31\textwidth]{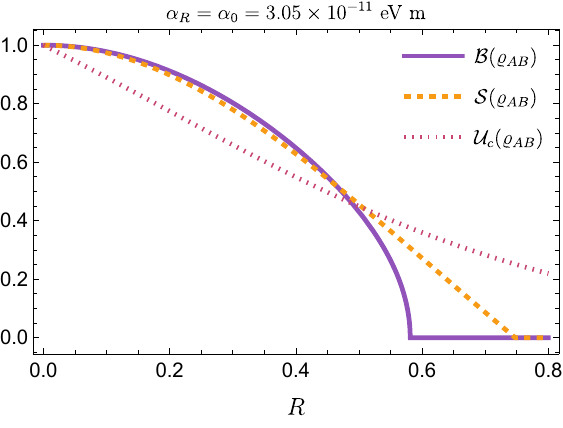}
		\label{1e}}
	\caption{Dynamics of ${\cal B}(\varrho_{AB})$ \ref{1a}, ${\cal S}(\varrho_{AB})$ \ref{1b}, and ${\cal U}_c(\varrho_{AB})$ \ref{1c} as a function of $R= \frac{|r-r'|}{r_0}$, with $r_0=7.14\times10^{-7}\text{cm}$  in two scenarios: without and with RSOI. A comparison of these three metrics in both scenarios \ref{1d}-\ref{1e}.}
	\label{fig1}
\end{figure*}

We examine the behavior of quantum resources in the Bi/Ag(111) system. To isolate RSOI-driven effects, we exclude Coulomb interactions and focus our analysis on spin-texture structure and exchange-hole suppression. We probe how RSOI strength ($\alpha_R$) and electron separation ($R$) modulate Bell nonlocality ${\cal B}(\varrho_{AB})$ (Eq. (\ref{eq31})), uncertainty-induced nonlocality ${\cal U}_c(\varrho_{AB})$ (Eq. (\ref{eq:UIN})), and quantum steering ${\cal S}(\varrho_{AB})$ (Eq. (\ref{eq33})). The Bi/Ag(111) system, with $m^* = 0.35~m_0$ and notable for its strong RSOI ($\alpha_0 = 3.05\times10^{-11}$ eV~m) \cite{aranzadi2023quantum} arising from surface alloying and structural corrugation \cite{bian2013origin, bian2012illuminating}, exhibits extreme Rashba spin splitting energies exceeding $200$ meV \cite{ast2007giant} in engineered 2DEG heterostructures and possesses exceptional spintronic properties, characterized by an electron density  $n = 6.25 \times 10^{11}\,\mathrm{cm^{-2}}$ and a characteristic radius $r_0 = (\sqrt{\pi n})^{-1} = 7.14 \times 10^{-7}\,\mathrm{cm}$. To investigate steerability and quantum nonlocality, we analyze Bell nonlocality $\mathcal{B}(\varrho_{AB})$ (Fig.~\ref{1a}), quantum steering $\mathcal{S}(\varrho_{AB})$ (Fig.~\ref{1b}), and uncertainty-induced nonlocality $\mathcal{U}_c(\varrho_{AB})$ (Fig.~\ref{1c}) as functions of electron separation $R$, comparing scenarios with and without RSOI. A comparative analysis (Figs.~\ref{1d},~\ref{1e}) reveals how RSOI modulates these resources via spin-texture asymmetry and exchange-hole suppression.

Fig.~\ref{fig1} shows the combined effect of RSOI and inter-electron separation $R$ on all three metrics in the Bi/Ag(111) system. Both $\mathcal{B}(\varrho_{AB})$ and $\mathcal{S}(\varrho_{AB})$ decay monotonically with $R$. They vanish at critical distances $R_c^{(B)}$ and $R_c^{(S)}$ that shift when RSOI is introduced: steering drops from $R_c^{(S)} = 0.78$ to $0.75$, and Bell nonlocality from $R_c^{(B)} = 0.61$ to $0.58$, upon switching on $\alpha_0 = 3.05\times10^{-11}$~eV\,m. Two mechanisms drive this suppression. First, exchange-hole reduction: RSOI disrupts the spatial overlap of same-spin electrons, weakening Pauli-exclusion-driven correlations in the two-particle density matrix~\cite{saitoh2006conversion}. Second, spin-momentum locking couples spin orientation to momentum direction, modifying the energy-branch occupations~\cite{bercioux2015quantum}. The numbers are unambiguous. At $R = 0.5$, $\mathcal{B}(\varrho_{AB})$ falls from 0.49 to 0.43 and $\mathcal{S}(\varrho_{AB})$ from 0.50 to 0.46 when RSOI is present. $\mathcal{U}_c(\varrho_{AB})$~\cite{wu2014uncertainty} behaves differently. It persists beyond $R_c^{(S)}$, retaining values of 0.24 (no RSOI) and 0.22 (with RSOI) at $R = 0.8$. This persistence reflects the structure of Eq.~(7): with $\mathbf{v} = 0$, $\mathcal{U}_c$ is determined by $n_{\min}(\mathbf{N})$, which depends on $\Gamma_2$ and therefore on spin-texture asymmetry rather than on direct spatial wavefunction overlap. Figures~(Figs.~\ref{1d},~\ref{1e}) compare all three metrics with and without RSOI, showing that $\alpha_0 = 3.05 \times 10^{-11}$~eV~m reduces the critical distances $R_c^{(B)}$ and $R_c^{(S)}$ while leaving $\mathcal{U}_c$ comparatively robust, reduced only mildly (from 0.24 to 0.22 at $R=0.8$) relative to the qualitative suppression of $\mathcal{B}$ and $\mathcal{S}$. \\

\begin{figure*}[t]
	\centering
	\subfloat[]{%
		\includegraphics[width=.31\textwidth]{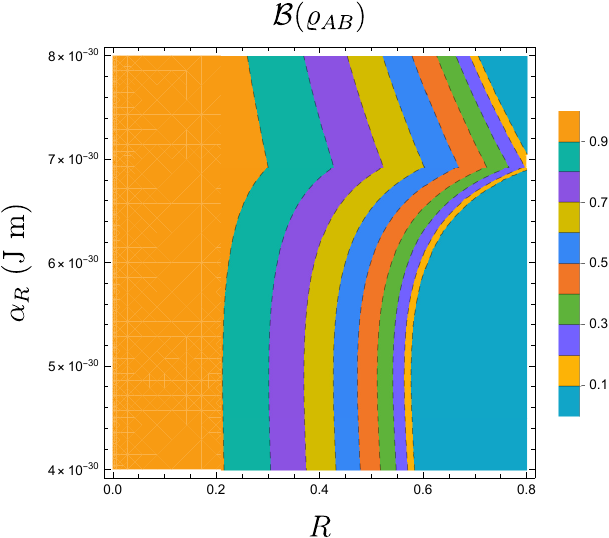}
		\label{2a}}
	\hfill
	\subfloat[]{%
		\includegraphics[width=.31\textwidth]{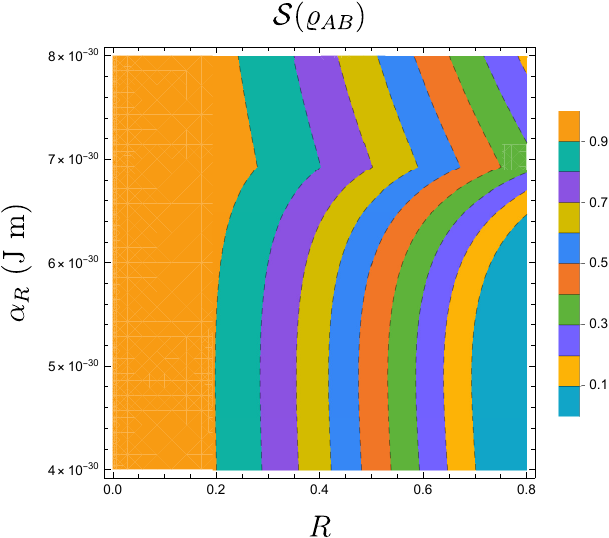}
		\label{2b}}
	\hfill
	\subfloat[]{%
		\includegraphics[width=.31\textwidth]{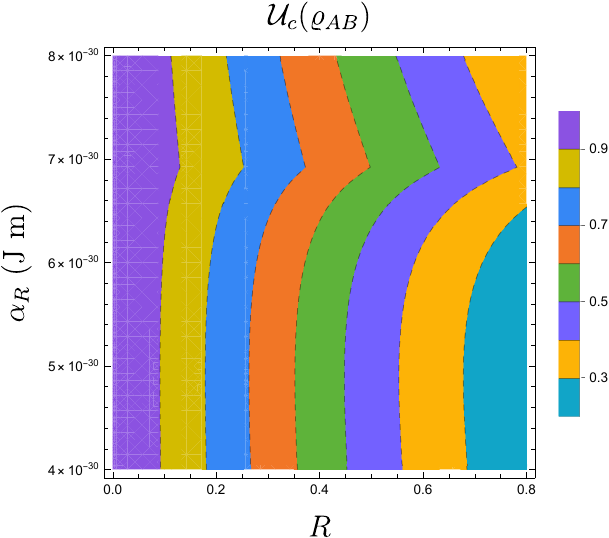}
		\label{2c}}\\
	\subfloat[]{%
		\includegraphics[width=.31\textwidth]{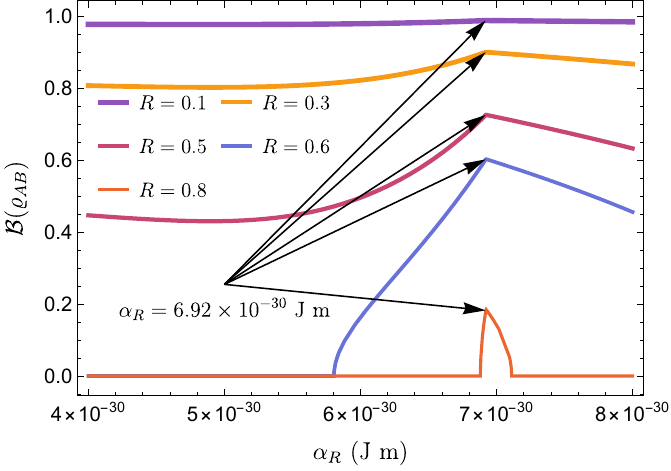}
		\label{2d}}
	\hfill
	\subfloat[]{%
		\includegraphics[width=.31\textwidth]{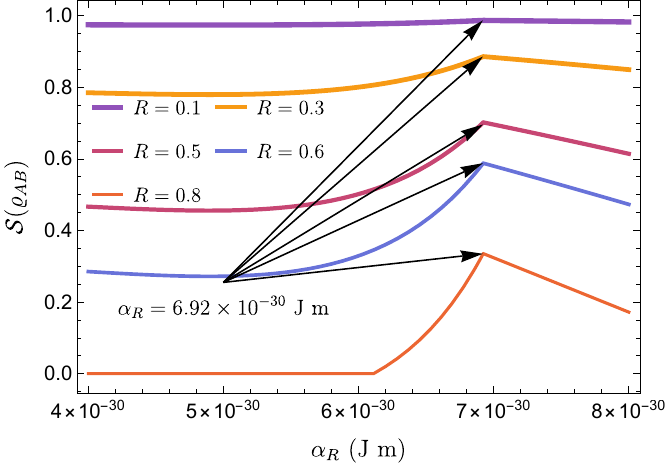}
		\label{2e}}
	\hfill
	\subfloat[]{%
		\includegraphics[width=.31\textwidth]{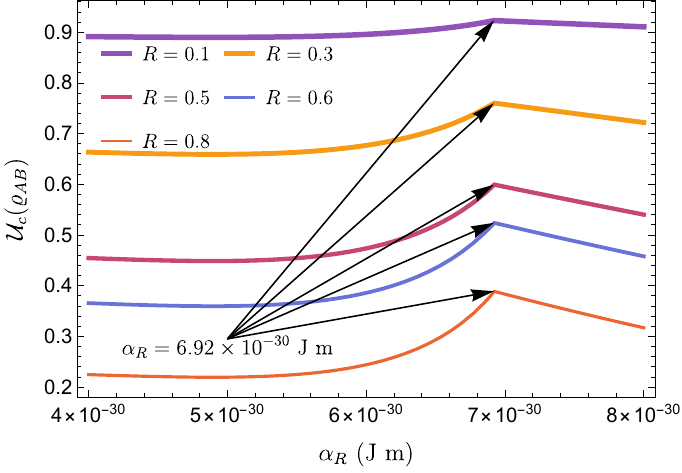}
		\label{2f}}
	\caption{Dynamics of ${\cal B}(\varrho_{AB})$, ${\cal S}(\varrho_{AB})$, and ${\cal U}_c(\varrho_{AB})$ as functions of $R$ and $\alpha_R$ \ref{2a}-\ref{2b}-\ref{2c} and as functions of $\alpha_R$, where $\alpha_0\le\alpha_R\le2\alpha_0$ with $\alpha_0 = 3.05\times10^{-11}$ eV~m, for various fixed values of $R$ \ref{2d}-\ref{2e}-\ref{2f}.}
	\label{fig2}
\end{figure*}

We now explore the impact of varying RSOI ($\alpha_R$; multiple values of $\alpha_0$) on the dynamic behavior of quantum steering, Bell nonlocality, and UIN. To illustrate this, we analyze $\mathcal{B}(\varrho_{AB})$, $\mathcal{S}(\varrho_{AB})$, and $\mathcal{U}_c(\varrho_{AB})$ as functions of $R$ for different values of $\alpha_R$, as shown in Figs. \ref{2a}, \ref{2b}, and \ref{2c}. Next, by keeping $R$ constant, we plot these metrics against $\alpha_R$, as presented in Figs. \ref{2d}, \ref{2e}, and \ref{2f}.

Figs.~\ref{2a}-\ref{2c} display $\mathcal{B}$, $\mathcal{S}$, and $\mathcal{U}_c$ as functions of both $R$ and $\alpha_R$, with $\alpha_R$ varied over $[\alpha_0,\,2\alpha_0]$. For numerical plotting we express $\alpha_R$ in SI units (J\,m); e.g., the optimal value $\alpha_R = 4.32\times10^{-11}$~eV~m corresponds to $6.92\times10^{-30}$~J~m.  As shown in Fig.~\ref{fig1}, introducing RSOI ($\alpha_R = \alpha_0 = 3.05\times10^{-11}$ eV~m) disrupts spatial electron correlations, suppresses the exchange hole, and reshapes the energy and momentum distributions. RSOI plays a dual role: correlations are enhanced or suppressed depending on $\alpha_R$, with the outcome governed by competing mechanisms. At baseline $\alpha_R$, exchange-hole suppression ($\Gamma_1$ in Eq.~(\ref{go})) attenuates Pauli exclusion-driven correlations, while elevated $\alpha_R$ amplifies spin-texture asymmetry ($\Gamma_2$ in Eq.~(\ref{ga})), reviving correlations via skew information. Figs.~\ref{2d}-\ref{2f} reveal that increasing $R$ consistently degrades ${\cal S}(\varrho_{AB})$ and ${\cal B}(\varrho_{AB})$, but elevating $\alpha_R$ from a relatively small value ($4\times 10^{-30}$ J~m) to $8\times 10^{-30}$ J~m (peaking at $\alpha_R = 6.92 \times 10^{-30}$ J~m) temporarily raises these metrics before they eventually decline. ${\cal U}_c(\varrho_{AB})$ retains $\mathcal{U}_c(\varrho_{AB}) \sim 0.22$ even at $R = 0.8$. Unlike Bell nonlocality or steering, ${\cal U}_c(\varrho_{AB})$ \cite{wu2014uncertainty} quantifies residual coherence from asymmetric spin textures ($\Gamma_2$), which are insensitive to direct wavefunction overlap. The electric-field tunability of $\alpha_R$ via gate voltages, demonstrated in related 2DEG systems~\cite{rendell2022gate}, suggests that the predicted enhancement of the considered quantum resources at the optimal $\alpha_R$ could in principle be accessed experimentally. This tunability is advantageous in Bi/Ag(111) heterostructures, where the system's strong intrinsic RSOI allows switching between suppression and recovery of correlations by adjusting $\alpha_R$. These results show that the optimal $\alpha_R \approx 4.32\times10^{-11}$~eV\,m recovers the critical distances to values close to those of the RSOI-free case, partially compensating the separation-induced degradation of Bell nonlocality and steering. \\

Finally, we investigate the effect of spatial separation between electrons $R $ on the quantum resources within the system for fixed values of $ \alpha_R $.
\begin{figure*}[t]
	\centering
	\subfloat[]{%
		\includegraphics[width=.31\textwidth]{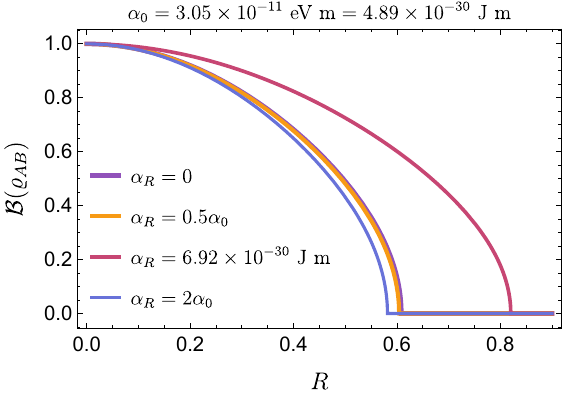}
		\label{3a}}
	\hfill
	\subfloat[]{%
		\includegraphics[width=.31\textwidth]{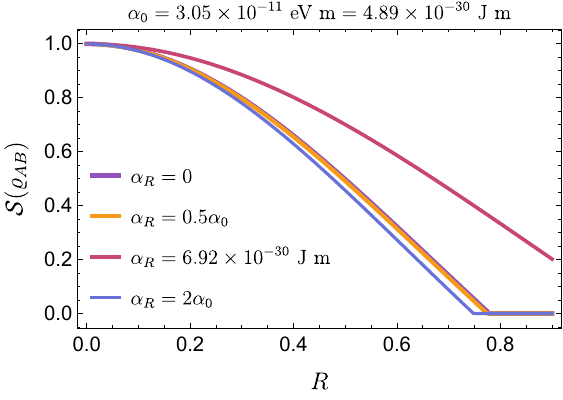}
		\label{3b}}
	\hfill
	\subfloat[]{%
		\includegraphics[width=.31\textwidth]{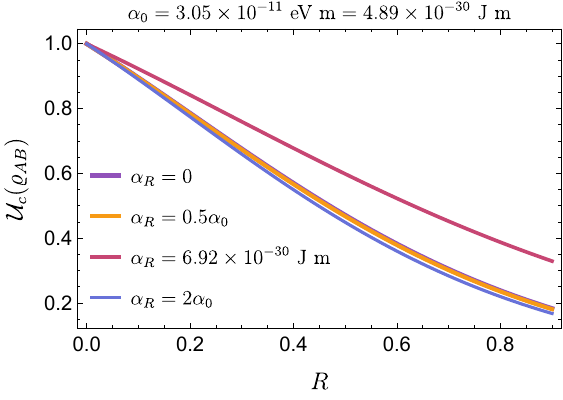}
		\label{3c}}
	\caption{Dynamics of ${\cal B}(\varrho_{AB})$ \ref{3a}, ${\cal S}(\varrho_{AB})$ \ref{3b}, and ${\cal U}_c(\varrho_{AB})$ \ref{3c} in terms of $R$ for different multiple values of $\alpha_0$ ($\alpha_R$) where $\alpha_0 = 3.05\times10^{-11}$ eV~m.}
	\label{fig3}
\end{figure*}

Figs.~\ref{3a}, \ref{3b}, and \ref{3c} show the evolution of Bell nonlocality $\mathcal{B}(\varrho_{AB})$, quantum steering $\mathcal{S}(\varrho_{AB})$, and uncertainty-induced nonlocality $\mathcal{U}_c(\varrho_{AB})$ in a 2DEG system as functions of electron separation $R$ for varying RSOI strengths $\alpha_R$. These results align with trends observed in Fig.~\ref{fig1}, where all metrics peak at $R \to 0$, reflecting maximal quantum correlations due to overlapping wavefunctions, and decay monotonically with increasing $R$. A sharp transition occurs at critical distances $R_c^{(B)}$ and $R_c^{(S)}$, beyond which $\mathcal{B}(\varrho_{AB})$ and $\mathcal{S}(\varrho_{AB})$ vanish, while UIN remains nonzero throughout, including at $R > 0.8$. The critical distance $R_c^{(B)}$ and $R_c^{(S)}$ for Bell nonlocality (and steering) decrease with increasing $\alpha_R$, shifting from $R^{(B)}_c = 0.6~(R^{(S)}_c = 0.77)$ at $\alpha_R = 0.5\alpha_0$ to $R^{(B)}_c = 0.58~(R^{(S)}_c = 0.75)$ at $\alpha_R = 2\alpha_0$. This suppression arises from the dual role of RSOI: at low $\alpha_R$, spin-momentum locking enhances correlations at small $R$ by coupling spin orientation $(\hat{\mathbf{s}})$ to electron momentum  $\hat{\mathbf{s}} \perp \mathbf{k}$, while at elevated $\alpha_R$, enhanced spin splitting suppresses the exchange hole, accelerating correlation decay by reducing Pauli exclusion-driven wavefunction overlap. UIN remains nonzero throughout: for example, $\mathcal{U}_c(\varrho_{AB}) = 0.18$ at $R = 0.9$ and $\alpha_R = 0.5\alpha_0$, a persistence attributed to asymmetric spin-texture correlations governed by $\Gamma_2$ (Eq.~\eqref{ga}). Unlike Bell nonlocality or steering, UIN quantifies skew information, which captures coherence from RSOI-induced antiparallel-spin correlations that are more resilient to spatial separation than those probed by Bell inequalities or steering witnesses.

Taken together, these results show that RSOI both suppresses and generates quantum correlations depending on the parameter regime: exchange-hole reduction dominates at baseline $\alpha_R$, while spin-texture asymmetry ($\Gamma_2$) prevails near the optimal $\alpha_R \approx 4.32\times10^{-11}$~eV~m.

\section{\label{sec:level5}Conclusion}

In this paper, we have examined a pair of non-interacting, delocalized electrons in a 2DEG under Rashba-type SOI, working at zero temperature and neglecting decoherence. This scope complements recent studies of the same resource triad under thermal and dephasing effects in localized spin platforms~\cite{oumennana2025thermal, bougouffa2026magnetic, mohamed2026enhancing}, isolating the purely coherent, separation and coupling-driven dynamics of Bell nonlocality, steering, and uncertainty-induced nonlocality in a genuinely delocalized many-body setting. We focus in particular on the Bi/Ag(111) system, which exhibits exceptionally strong RSOI ($\alpha_0 =3.05\times10^{-11}$~eV\,m) due to surface alloying and structural corrugation~\cite{aranzadi2023quantum, ast2007giant, bian2013origin, bian2012illuminating}, with reported Rashba splitting energies exceeding 200~meV~ \cite{ast2007giant}. The system has an electron density $n = 6.25\times10^{11}$~cm$^{-2}$ and characteristic radius $r_0 = (\sqrt{\pi n})^{-1} = 7.14\times10^{-7}$~cm. We probe how RSOI strength ($\alpha_R$) and electron separation ($R$) modulate quantum steering (${\cal S}(\varrho_{AB})$), Bell nonlocality (${\cal B}(\varrho_{AB})$), and uncertainty-induced nonlocality (${\cal U}_c(\varrho_{AB})$) and we show theoretically that quantum resources in this class of system can be preserved and enhanced by carefully tuning the RSOI strength. The results show that increasing the inter-electron distance $R$ consistently degrades quantum resources, whereas a Rashba coupling strength of $\alpha_R=4.32\times10^{-11}$ eV~m (corresponding to $\alpha_R = 6.92 \times 10^{-30}$ J~m) achieves a maximum of these metrics before they eventually decline. The uncertainty-induced nonlocality ${\cal U}_c(\varrho_{AB})$ persists even at $R=0.8$, retaining nonzero values. This makes it more resilient to spatial separation than quantum steering and Bell nonlocality. Finally, we note that RSOI both suppresses and enhances correlations relative to the no-RSOI limit, depending on its strength. At baseline $\alpha_R$, exchange-hole suppression attenuates Pauli exclusion and reduces quantum correlations, while tuning $\alpha_R$ to specific optimal values (around $4.32\times10^{-11}$ eV~m) restores them.

\bibliography{Refs}
\bibliographystyle{ieeetr}

\end{document}